\documentclass[]{pasj01}
\draft

\begin{document} 
\Received{}
\Accepted{}

\title{Accretion environments of active galactic nuclei}

\author{Hajime \textsc{Inoue}\altaffilmark{1}}%
\altaffiltext{1}{Institute of Space and Astronautical Science, Japan Aerospace Exploration Agency, 3-1-1 Yoshinodai, Chuo-ku, Sagamihara, Kanagawa 252-5210, Japan}
\email{inoue-ha@msc.biglobe.ne.jp}


\maketitle

\begin{abstract}
We study accretion environments of active galactic nuclei when a super-massive black hole wanders in a circum-nuclear region and passes through an interstellar medium there.
It is expected that a Bondi-Hoyle-Lyttleton type accretion of the interstellar matter takes place and an accretion stream of matter trapped by the black hole gravitational field appears from a tail shock region.
Since the trapped matter is likely to have a certain amount of specific angular momentum, the accretion stream eventually forms an accretion ring around the black hole.
According to the recent study, the accretion ring consists of a thick envelope and a thin core, and angular momenta are transfered from the inner side facing to the black hole to the opposite side respectively in the envelope and the core. As a result, a thick accretion flow and a thick excretion flow extend from the envelope, and a thin accretion disk and a thin excretion disk do from the core. 
The thin excretion disk is predicted to terminate at some distance forming an excretion ring, while the thick excretion flow is considered to become a super-sonic wind flowing to the infinity.
The thick excretion flow from the accretion ring is expected to interact with the accretion stream toward the accretion ring and to be collimated to bi-polar cones.
These pictures provide a likely guide line to interpret the overall accretion environments suggested from observations.

\end{abstract}

\section{Introduction}
It is now widely accepted that activities of active galactic nuclei (AGNs) are results of mass accretion on to a super-massive black hole at the center of a galaxy, and it is strongly supported by presence of several observational similarities between AGNs and black hole X-ray binaries (e.g. Inoue 2021c for the review).

Studies of the accretion environments done so far have been based on a picture that matter in a circum-nuclear region with 10 - 100 pc scales continuously inflow through a dusty torus to a sub-pc region, connecting to an accretion disk on to the black hole sitting at the center.
This picture, however, faces with some difficult issues such as the angular momentum barrier, or the star burst barrier (see e.g. Jogee 2006; Alexander \& Hickox 2012, for the review).
The connections of the dusty torus, the broad line region and the accretion disk are also still open questions (see e.g. Czerny 2019 for the review).

An interesting possibility to resolve the difficulties could be that 
the black hole rather comes to the circum-nuclear region than stays at the center, and passes through an interstellar medium there, inducing the Bondi-Hoyle-Lyttleton (BHL) type accretion (e.g. Edgar 2004 for the review) as often exhibited in high mass X-ray binaries (see e.g. Davidson \& Ostriker 1973).
Recently, Inoue (2021a) proposes a possible mechanism for the super-massive black hole to wander in a circum-nuclear region at a several 10 pc distance.

The BHL-type accretion flows had been theoretically studied in many papers and presence of radial and tangential (`flip-flop') instabilities had been pointed out before $\sim$2000 (see Edger 2004 for the review and references therein).
Recent three-dimensional simulations of the BHL-type accretion flows with higher spatial resolution by Blondin and Raymer (2012), however, showed that a radial oscillation causing an accretion rate modulation with amplitude of 20\% could arise but that the flow remained highly axis-symmetric with only negligible accretion of angular momentum.

Instabilities associated with the ionization front in the BHL-type accretion flows were investigated, considering radiative feedback effects, in two-dimensional axis-symmetric simulations by Park and Ricotti (2013) and in three-dimensional simulations by Sugimura and Ricotte (2020).
The results indicate that unstable situations appear in fairly wide parameter ranges of the flow.
These instabilities under the effects of the radiation from the central engine are discussed to be suppressed in the present study.

Hereafter, we study accretion environments in the case of a passage of a wandering super-massive black hole through an interstellar medium in the circum-nuclear region, where the BHL accretion flow is likely to take place.

Average properties of energy spectra and power density spectra observed from AGNs are similar to those of the black hole binaries (see e.g. Inoue 2021c), and are consistent with a picture in which a steady accretion disk around the central black hole is responsible for the AGN activities.
This requires that the BHL accretion flow should be stable within a scope in which snapshot observations of many AGNs exhibit no large deviation from the average properties.

The presence of the accretion disk needs a situation in which a certain amount of angular momentum should be carried to the outermost part of the accretion disk.
In such a situation, the matter trapped by the potential of the black hole in the BHL accretion flow is considered to once form a ring at the Keplerian circular orbit determined by the specific angular momentum of the matter inflowing to the ring, before extension of an accretion disk.
Inoue (2021b) studies properties of the ring (called the accretion ring) and shows that the accretion ring has a thick envelope and a thin core, and generates a two-layer accretion flow toward the central black hole in which a thin accretion disk is sandwiched by a thick accretion flow.
At the same time, it is predicted that another two-layer flow consisting of 
 a thin excretion disk and a thick excretion flow extends outward from the accretion ring and that the thin excretion disk terminates at a certain distance, while the thick excretion flow eventually becomes a super-sonic wind to the infinity.
If we apply these arguments on the accretion ring to the present case, excretion flows can be expected to appear in the AGN environments and are likely to interact with the BHL accretion flow.

The unified model for AGNs based on observations (e.g. Antonucci 1993) indicates that radiation from the central engine is distributed preferentaially to the polar directions, and the recent observations with IR and submillimeter interferometers reveal that matter is also out-flowing to the polar directions in association with the radiation cones (see e.g,H\"{o}nig 2019 and references therein).
We discuss that these situations can be results of the interactions between the accretion flow to the accretion ring and the excretion flow from it.
It is also suggested that the BHL accretion flows could take place in a region along the equatorial plane, sandwiched by the two polar cones with fairly large opening angle, where the radiation from the central engine is almost blocked by the out-flowing matter in the excretion flow.
In such a circumstance, the instabilities due to the effects of the radiation on the BHL accretion flow is expected to be sufficiently suppressed.

\begin{figure}
 \begin{center}
  \includegraphics[width=10cm]{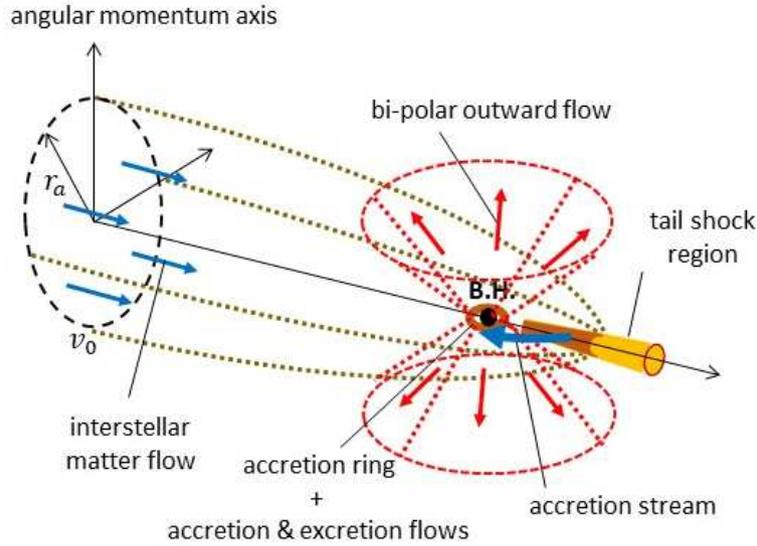}
 \end{center}
\caption{Schematic diagram of the Bondi-Hoyle-Lyttleton type accretion of interstellar matter by a super-massive black hole wandering in a circum-nuclear region.
The interstellar matter passes by the black hole (B.H.) with an incident velocity, $v_{0}$,  and collides with one another behind B.H., forming a tail shock region.  The matter within the accretion radius, $r_{\rm a}$, is trapped by the gravity of B.H., inflows through an accretion stream toward B.H. and eventually forms an accretion ring around B.H. in a plane determined by their angular momentum axis.  The accretion ring extends thin and thick excretion flows as well as thin and thick accretion flows.  The thick excretion flow turns to bi-polar flows by its interaction of the accretion stream.  The central region around the black hole (B.H.) is expanded in figure \ref{fig:ExcretionFlow}.}
\label{fig:OverviewPicture}
\end{figure}

\begin{figure}
 \begin{center}
  \includegraphics[width=10cm]{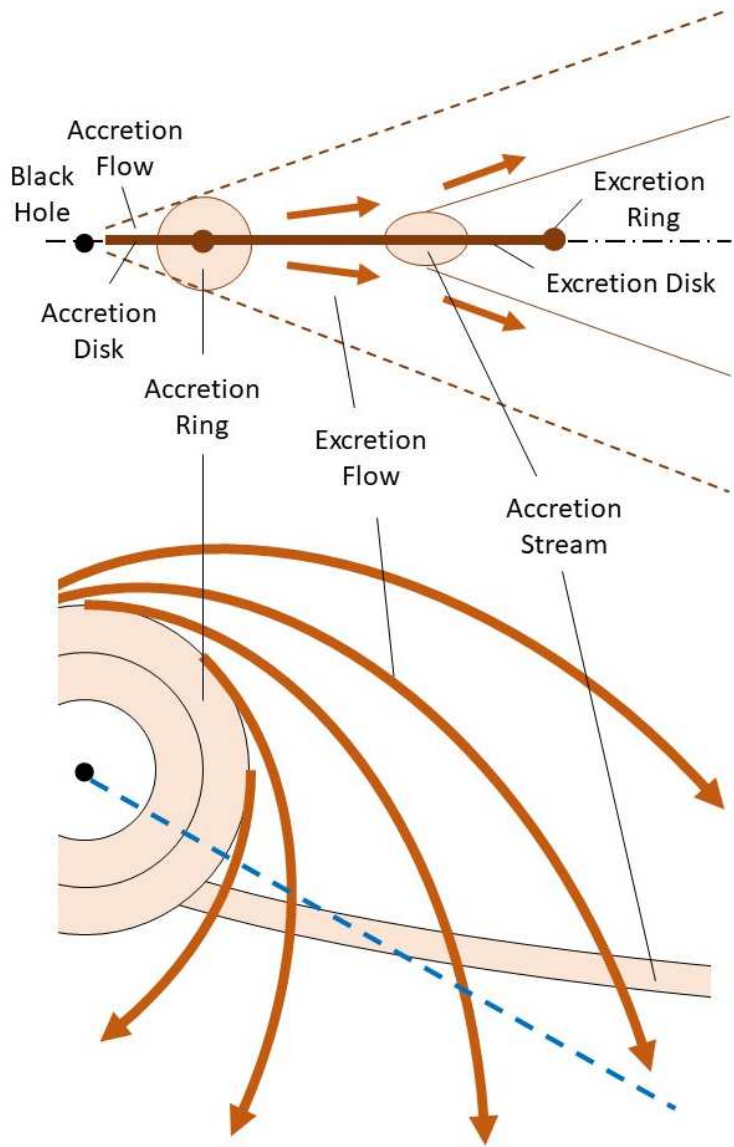}
 \end{center}
\caption{(Bottom) a schematic top view showing geometrical relations between flow lines of the excretion flow from the accretion ring and the accretion stream and (Top) a cross section along the dashed line in the bottom diagram indicating configurations of several flows discussed in this study.}
\label{fig:ExcretionFlow}
\end{figure}

Here, we briefly introduce several steps for the interstellar matter to experience after it is captured by the gravitational field of the black hole, which are considered in this study (see figures \ref{fig:OverviewPicture} and \ref{fig:ExcretionFlow} for the reference):

In the rest frame of the black hole, the interstellar matter is supposed to flow in parallel from the infinity toward the black hole.
As briefly discussed above, the interstellar matter first interacts with the out-flowing matter from the accretion ring and is forced to bypass the excretion flows.
Nevertheless, most of the matter is expected to eventually gather and collide with one another behind the black hole.  
This flow of the interstellar matter is called "the ISM flow".
A tail shock region appears on the back-side of the black hole, in which the velocity component perpendicular to the tail shock surface is converted to the thermal motion and its energy is assumed to be radiated away via thermal emission.
As a result of the partial removal of the kinetic energy in the tail shock region, the interstellar matter having the impact parameter less than the accretion radius defined later in equation (\ref{eqn:r_a}) is trapped by the black hole gravitational field.  It starts flowing toward the black hole, and that flow is designated "the accretion stream".

Since it would be natural for the matter flowing from the tail shock region into the accretion stream to have a certain amount of specific angular momentum, the stream is expected to eventually form the accretion ring at the Keplerian circular orbit determined by the specific angular momentum.
The accretion ring has a thick envelope around a thin core and generates a two-layer accretion flow toward the central black hole in which a thin accretion disk is sandwiched by a thick accretion flow.
At the same time, another two-layer flow consisting of 
 a thin excretion disk and a thick excretion flow extends outward from the accretion ring and that the thin excretion disk terminates at a circular orbit with a radius 4 times the accretion ring radius, forming another ring which we call ``the excretion ring", while the thick excretion flow eventually becomes a super-sonic wind to the infinity.

\begin{figure}
 \begin{center}
  \includegraphics[width=14cm]{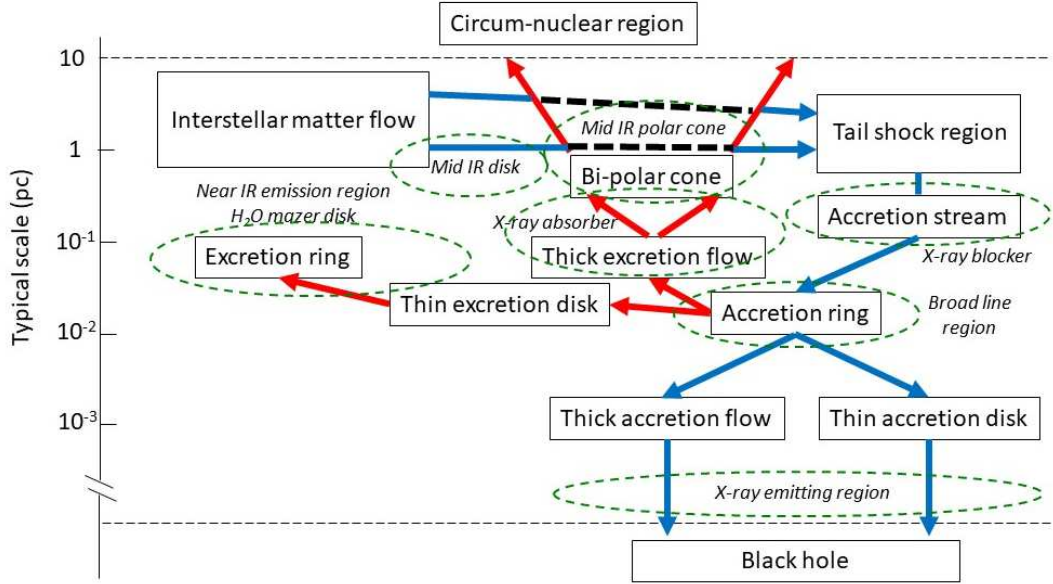}
 \end{center}
\caption{Block diagram of elements studied in this study and matter-flow-directions between them.  Candidate places for the observed elements are also indicated with dashed ellipses and italic letters.  The left vertical line indicates the typical scales for $M = 10^{7} M_{\odot}$.  For details, see the text.}
\label{fig:FlowChart}
\end{figure}

These components are listed in a form of block diagram to summarize the present study, in figure \ref{fig:FlowChart}.

We study properties of each of such elements: the ISM flow, the tail shock region, the stream, the accretion ring, the thin excretion disk and the thick excretion flow, discuss  some interactions between them, and compare the results with observations, hereafter.

\section{Accretion environments of AGNs}

\subsection{Accretion of interstellar matter by a wandering black hole}
We first assume that a super-massive black hole wanders in a galactic nuclear region and often come to a circum-nuclear region at several 10 pc from the nuclear center.
We then consider a situation that the black hole passes through an interstellar medium in the circum-nuclear region and a BHL type accretion of the matter takes place on to the black hole, inducing AGN activities.

If interstellar matter in a galactic nuclear region is approaching with velocity, $v_{0}$, to a super-massive black hole with mass $M$,
the accretion radius, $r_{\rm a}$, the largest impact parameter within which the matter is captured by the black hole gravitational field, can be calculated as
\begin{eqnarray}
r_{\rm a} &=& \frac{2GM}{v_{0}^{2}} \nonumber \\
&=& 8.6 \left(\frac{M}{10^{7} M_{\odot}}\right) \left(\frac{v_{0}}{10^{7} \mbox{cm s}^{-1}}\right)^{-2} \mbox{ pc}. 
\label{eqn:r_a}
\end{eqnarray}
Adapting $v_{0} \sim 10^{7}$ cm s$^{-1}$ as the typical velocity  of the circum-nuclear matter, the accretion radius would be of the order of 10 pc for the black hole mass of $10^{7} \sim 10^{8} M_{\odot}$.

The mass inflow rate to the black hole field, $\dot{M}_{0}$, and the bolometric luminosity as a result of the final mass accretion on to the black hole, $L$, are approximately given as
\begin{eqnarray}
\dot{M}_{0} &\simeq& \pi r_{\rm a}^{2} n_{\rm is} m_{\rm p} v_{0} \nonumber \\
&\simeq& \frac{\eta c \sigma_{\rm T} n_{\rm is} GM}{2 v_{0}^{3}} \dot{M}_{\rm E} \nonumber \\ &=& 1.3 \times 10^{-1} \left(\frac{\eta}{0.1}\right) \left(\frac{n_{\rm is}}{10^{2} \mbox{ cm}^{-3}}\right) \left(\frac{v_{0}}{10^{7} \mbox{ cm s}^{-1}}\right)^{-3} \left(\frac{M}{10^{7} M_{\odot}}\right) \dot{M}_{\rm E},
\label{eqn:Mdot_0}
\end{eqnarray}
and 
\begin{eqnarray}
L &\simeq& \eta \frac{\dot{M}_{0}}{2} c^{2} \nonumber \\
&\simeq& \frac{ \eta 2 \pi (GM)^{2} n_{\rm is} m_{\rm p} c^{2}}{v_{0}^{3}} \nonumber \\
&=& 1.7 \times 10^{44} \left(\frac{\eta}{0.1}\right) \left(\frac{v_{0}}{10^{7}\mbox{ cm s}^{-1}}\right)^{-3} \left(\frac{n_{\rm is}}{10^{2}\; \rm{cm}^{-3}}\right)  \left(\frac{M}{10^{7}M_{\odot}}\right)^{2} \; \textrm{erg s}^{-1},
\label{eqn:L_X}
\end{eqnarray}
where $n_{\rm is}$ is the number density of the interstellar matter in the nuclear region,
$\eta$ is the energy conversion efficiency of the accretion matter on to the black hole and $m_{\rm p}$, $c$ and $\sigma_{\rm T}$ are the proton mass, the light velocity and the Thompson scattering cross section, respectively.
In the estimation of $L$, it is assumed that a half of $\dot{M}_{0}$ is finally accreted by the black hole, which is discussed later.
$\dot{M}_{\rm E}$ is the mass inflow rate eventually yielding the Eddington luminosity defined as
\begin{equation}
\dot{M}_{\rm E} = \frac{2}{\eta c^{2}} \frac{4\pi c GM m_{\rm p}}{\sigma_{\rm T}}.
\label{eqn:Mdot_E}
\end{equation}
If we adopt $n_{\rm is} \simeq 10^{2}$ cm$^{-3}$ from the average number density of the molecular gas in the inner nuclear bulge of our Galaxy obtained by Launhardt, Zylka \& Mezger (2002), $v_{0} \simeq 10^{7}$ cm s$^{-1}$, 
and $\eta \simeq 0.1$, $L$ estimated from equation (\ref{eqn:L_X}) roughly agrees to the observed values taking account of ambiguities in the estimation.

\subsection{Tail shock region}\label{TailShockRegion}
The matter passing by the black hole should collide with one another behind the black hole and form a tail shock region there.
The temperature of the tail shock region, $T_{\rm ts}$, is approximately estimated as $T_{\rm ts} \sim (m_{\rm p}/k) v_{0}^{2}/10 \simeq 1 \times 10^{5}$ K for $v_{0} = 10^{7}$ cm s$^{-1}$, where $k$ is the Boltzmann constant.  Here, we assume that the matter is fully ionized hydrogen gas with a specific heat ratio of 5/3.
If we set the number density of the gas behind the shock to be 10 $n_{\rm is} \sim 10^{3}$ cm$^{-3}$, considering the flow convergence and the shock compression, and $T_{\rm ts} \simeq 10^{5}$ K, the cooling time, $t_{\rm c, \; ts} \simeq 3 k T_{\rm ts}/(n_{\rm is} \Lambda(T_{\rm ts})$, is estimated to be $4 \times 10^{7}$ s for $\Lambda(T_{\rm ts}) \sim 1 \times 10^{-21}$ erg cm$^{3}$ s$^{-1}$ (see e.g. Sutherland \& Dopita 1993).
This cooling time is much shorter than the typical flow time, $r_{a}/v_{0} \sim 3 \times 10^{12}$ s for $M \simeq 10^{7} M_{\odot}$ and $v_{0} \simeq 10^{7}$ cm s$^{-1}$ and thus the cooling should effectively happen.
The luminosity of the tail shock region, $L_{\rm ts}$, is roughly given with the help of equation (\ref{eqn:L_X}) as 
$L_{\rm ts} \sim \dot{M}_{0} v_{0}^{2}/2 \simeq v_{0}^{2} L /(\eta c^{2})
 \sim 1 \times 10^{-6} L$ for $v_{0} = 10^{7}$ cm s$^{-1}$ and $\eta = 0.1$.
The short cooling time indicates that the thickness of tail shock region should be small compared with the radial size and thus its solid angle as viewed from the black hole would not be so large as to significantly contribute to X-ray obscuration.

The tail shock region would observationally be inconspicuous.

\subsection{Accretion stream}\label{stream}
Among the matter in the tail shock region, 
those having had the impact parameter no larger than $r_{\rm a}$ should start falling from the tail shock region and making the accretion stream towards the black hole.  
The matter flowing from the tail shock region is considered to have a certain amount of specific angular momentum but it could be so small that the stream is almost radial near the tail shock region.
As it approaches the black hole, the rotational component gradually increases and the accretion ring is eventually formed as discussed in the next subsection.

The mass flow rate in the accretion stream is $\dot{M}_{0}$ and is approximately expressed with its number density $n_{\rm st}$ and velocity $v_{\rm st}$ on the assumption that the stream has a spherical cross section with radius, $d$,  as
\begin{equation}
\dot{M}_{\rm 0} \simeq \pi d^{2} n_{\rm st} m_{\rm p} v_{\rm st}.
\label{eqn:Mdot_st}
\end{equation}
From this equation, the optical depth for the Compton scattering, $\tau_{\rm st}$, of the accretion stream across the cross section can roughly be estimated as
\begin{eqnarray}
\tau_{\rm st} &\simeq& 2n_{\rm st} d \sigma_{\rm T} \nonumber \\
&\simeq& \frac{2 \sqrt{2}}{\eta \delta} \frac{v_{0}}{c} \frac{\dot{M}_{0}}{\dot{M}_{\rm E}} \left(\frac{r}{r_{\rm k}}\right)^{-1/2} \left(\frac{d}{r}\right)^{-1} \nonumber \\
&\simeq& 3 \left( \frac{\eta}{0.1}\right)^{-1} \left(\frac{\delta}{10^{-1.5}}\right)^{-1} \left(\frac{v_{0}}{10^{7} \mbox{ cm s}^{-1}}\right) \left(\frac{d/r}{0.1}\right)^{-1} \left(\frac{\dot{M}_{0}}{\dot{M}_{\rm E}}\right) \left(\frac{r}{r_{\rm k}}\right)^{-1/2},
\label{eqn:tau_st}
\end{eqnarray}
where $v_{\rm st} = \sqrt{2GM/r}$ is assumed, and $\delta$ and $r_{\rm k}$ are introduced in equation (\ref{eqn:ell_a}) and equation (\ref{eqn:r_k}), respectively, in the next subsection. 
Adopting $\eta \sim 0.1$, $\delta \sim 10^{-1.5}$ and $v_{0} \sim 10^{7}$ cm s$^{-1}$, assuming $d \lesssim 0.1\ r$ and taking account of that the line of sight to the central X-ray source slantingly crosses the stream, the accretion stream near the accretion ring could be Compton thick against the X-rays on the line of sight, when $\dot{M}_{0} \gtrsim 0.1 \dot{M}_{\rm E}$.

\subsection{Accretion ring}\label{accretion_ring}
We can consider two possible origins for the angular momentum carried by the accretion stream.

One is a density gradient of the interstellar matter in a direction perpendicular to the flow direction before passing by the black hole.
The simple thought is that if more matter is accreted from one side than the other, the accreted matter could have a net non-zero angular momentum (e.g. Shapiro \& Lightman 1976).
However, theoretical study of such an accretion flow by Davies and Pringle (1980) reveals that the accreted angular momentum should be zero under the first order approximation. 
On the other hand, 3-dimensional simulations of the BHL accretion flows under the presence of the density gradients by Ruffert (1999) 
exhibit solutions having non-zero accretion rate of angular momentum.

The other is a Coriolis force in the black hole rest frame.
In the present study, the black hole is assumed to wander in a gravitational potential of the galactic nucleus by being attracted by nearby molecular clouds (Inoue 2021a), and is likely to have a rotational component in its motion. 
Thus, the black hole rest frame is expected to have an angular velocity relative to the inertial frame, and then the Coriolis force is considered to induce an angular momentum in the accretion stream.

In either case, a specific angular momentum, $\ell_{\rm a}$, carried by the accretion stream can be expressed as
\begin{equation}
\ell_{\rm a} = \delta r_{\rm a} v_{0},
\label{eqn:ell_a}
\end{equation}
where $\delta$ is a deviation factor from the normalization value $r_{\rm a} v_{0}$ and should be proportional to the degree of the density gradient and the strength   of the Coriolis force in the respective cases.

If the matter with the specific angular momentum as expressed above forms a ring with the Keplerian circular orbit, the radius, $r_{\rm k}$, is calculated as
\begin{eqnarray}
r_{\rm k} &=& \frac{\ell_{\rm a}^{2}}{GM} \nonumber \\
&\simeq& \frac{4 \delta^{2} GM}{v_{0}^{2}} \nonumber \\
&\simeq& 1.7 \times 10^{-2} \left(\frac{\delta}{10^{-1.5}}\right)^{2} \left(\frac{v_{0}}{10^{7}\; \textrm{cm s}^{-1}}\right)^{-2} \left(\frac{M}{10^{7}M_{\odot}}\right) \; \rm{pc}.
\label{eqn:r_k}
\end{eqnarray}
If we adopt $\delta \sim 10^{-1.5}$, $r_{\rm k}$ is estimated to be 0.02 pc for $M \simeq 10^{7} M_{\odot}$ and $v_{0} \simeq 10^{7}$ cm s$^{-1}$.
Referring to the luminosity in equation (\ref{eqn:L_X}) for the same $M$ and $v_{0}$ values, 
this radius roughly agrees to that of the broad line region determined by the optical reverberation mapping observations (e.g. Bentz et al. 2009).
Furthermore, we can get a relation of $r_{\rm k} \propto L^{1/2}$ by eliminating $M$ from $L \propto M^{2}$ in equation (\ref{eqn:L_X}) and $r_{\rm k} \propto M$ in equation (\ref{eqn:r_k}), unless $v_{0}$ or $\delta$ has a large dependency on $M$.
This could explain an observed proportionality between the radii of the broad line region and the square root of the optical luminosities (Bentz et al. 2009).

These discussions indicate that the broad line region corresponds to the ring which the matter from the accretion stream is expected to form.
In fact, some observed properties of the broad line regions are consistent with those of the accretion ring studied for X-ray binaries by Inoue (2021b) as discussed below.

Inoue (2021b) investigates properties of the accretion ring in an X-ray binary, where the accretion ring is defined as a sojourning place of matter inflowing from a companion star before turning to an accretion flow to a compact object.
It is discussed that the inflowing matter initially forms a hot and thick envelope along the ring.
Its temperature, $T_{\rm in}$, is approximately calculated from an equation as \begin{equation}
\frac{v_{\phi, \; \rm k}^{2}}{2} + \frac{5kT_{\rm in}}{m_{\rm p}} - \frac{GM}{r_{\rm k}} = \varepsilon_{0},
\label{eqn_EnEq-r_k}
\end{equation}
where $v_{\phi,\; \rm k} = \sqrt{GM/r_{\rm k}}$ is the rotational velocity of the ring and 
$\varepsilon_{0}$ (negative value) is the specific energy of the matter in the ring just after inflowing from the stream. 
Approximating $\varepsilon_{0} = 0$, we obtain 
\begin{eqnarray}
T_{\rm in} &\simeq& \frac{1}{10} \frac{m_{\rm p}GM}{kr_{\rm k}} \nonumber \\ &\simeq& 3.2 \times 10^{7} \left( \frac{v_{0}}{10^{7} \mbox{ cm s}^{-1}}\right)^{2} \left( \frac{\delta}{10^{-1.5}}\right)^{-2} \mbox{ K},
\label{eqn:T_in}
\end{eqnarray}
with the help of equation (\ref{eqn:r_k}), in the present case.

Two internal flows are expected to appear in the thick envelope.
One is a mass spreading flow bifurcating to a thick accretion flow and a thick excretion flow, as a result of the angular momentum transfer within the envelope.
The other is a cooling flow toward the envelope center governed by radiative cooling under the effect of X-ray irradiation. 
This cooling flow eventually forms a core around the center of the envelope, from which a thin accretion disk and a thin excretion disk spread out as a result of the angular momentum transfer there again.

The mass inflow with the rate $\dot{M}_{0}$ from the accretion stream is first bifurcated to the two internal flows in the envelope, the mass spreading flow with the rate $\dot{M}_{\rm tk}$ and the cooling flow with the rate $\dot{M}_{\rm tn}$, as given as
\begin{equation}
\dot{M}_{\rm tk} + \dot{M}_{\rm tn} = \dot{M}_{0}.
\label{eqn:Mdot_tk+Mdot_tn}
\end{equation}
Then, the mass spreading flow is evenly divided into the thick accretion flow with the rate $\dot{M}_{\rm tk,\; ac}$ and the thick excretion flow with the rate $\dot{M}_{\rm tk,\; ex}$, as expressed as
\begin{equation}
\dot{M}_{\rm tk,\; ac} = \dot{M}_{\rm tk,\; ex} = \frac{\dot{M}_{\rm tk}}{2}.
\label{eqn:Mdot_tk,ac=Mdot_tk,ex}
\end{equation}
In parallel, the cooling flow is evenly divided into the thin accretion disk with the rate, $\dot{M}_{\rm tn,\; ac}$ and the thin excretion disk with the rate $\dot{M}_{\rm tn,\; ex}$, as written as
\begin{equation}
\dot{M}_{\rm tn,\; ac} = \dot{M}_{\rm tn,\; ex} = \frac{\dot{M}_{\rm tn}}{2}.
\label{eqn:Mdot_tn,ac=Mdot_tn,ex}
\end{equation}

It is assumed above that the accretion rate and the excretion rate equal to each other for each of the thick and thin flows.
Here, we consider a situation in which the angular momentum carried into the accretion ring by the matter from the tail shock region is transfered from the inner half of the ring facing the black hole to the outer half through turbulent viscosity and the ring starts spreading out. 
Thus, the equal separation into the accretion flow and the excretion flow would be quite natural.  
In appendix, we estimate how much angular momentum and energy should be transfered from the accretion matter to the excretion matter in the general case with different accretion and excretion rates.
The results indicate that such an uneven case as $\dot{M}_{\rm tk,\; ex} \ll \dot{M}_{\rm tk,\; ac}$ or $\dot{M}_{\rm tn,\; ex} \ll \dot{M}_{\rm tn,\; ac}$ should be unreasonable.

Evaluating the timescales for the two internal flows, the mass spreading flow and the cooling flow, respectively and comparing them with each other, the properties of the internal flows are classified into the following three cases in terms of the intrinsic mass inflow rate from the stream, $\dot{M}_{0}$.
\begin{itemize}
\item Low inflow rate case
\end{itemize}
When $\dot{M}_{0}$ is much less than a boundary inflow rate, $\dot{M}_{1}$, the density of the ring-tube is too low for the matter to cool down and the ring-tube is kept thick with the initial temperature, $T_{\rm in}$.
Applying the calculation by Inoue (2021b) to the present case, $\dot{M}_{1}$ is given as
\begin{equation}
\dot{M}_{1} \simeq 9.1 \times 10^{-5} \left(\frac{\eta}{0.1}\right) \left( \frac{\alpha}{0.1} \right) \left( \frac{\Lambda}{\Lambda_{0}}\right)^{2} \left( \frac{\delta}{10^{-1.5}}\right)^{4} \left( \frac{v_{0}}{10^{7}}\right)^{-4} \left( \frac{T_{\rm in}}{3.2 \times 10^{7} \mbox{ K}}\right)^{3}\ \dot{M}_{\rm E},
\label{eqn:M_1}
\end{equation}
where $\Lambda$ is the cooling function and $\Lambda_{0} = 10^{-22.6}$ erg cm$^{3}$ s$^{-1}$ being the value for the temperature around 10$^{7} \sim 10^{8}$ K (Sutherland \& Dopita 1993).  The unit of $v_{0}$ is cm s$^{-1}$.

In this case, almost all the mass flow from the stream is converted to the thick flows from the envelope.

\begin{itemize}
\item Medium inflow rate case
\end{itemize}
When $\dot{M}_{0}$ is as large as or larger than $\dot{M}_{1}$ but much less than another boundary inflow rate, $\dot{M}_{2}$, as given below, the effect of the X-ray irradiation is not significant yet but two-layer accretion and excretion flows are expected to appear.
$\dot{M}_{2}$ is calculates as
\begin{equation}
\dot{M}_{2} \simeq 4.1 \times 10^{-2} \left( \frac{\eta}{0.1} \right)^{-1} \left( \frac{\Lambda}{\Lambda_{0}}\right)^{-1} \left( \frac{T_{\rm in}}{3.2 \times 10^{7} \mbox{ K}}\right)^{3}\ \dot{M}_{\rm E}.
\label{eqn:M_2}
\end{equation}

In this case, 
The two-layer flows have the comparable mass flow rate to each other but that in the thin flows tend to be dominant to the other as $\dot{M}_{0}$ increases.

\begin{itemize}
\item High accretion rate case
\end{itemize}
When $\dot{M}_{0}$ is as large as or larger than $\dot{M}_{2}$, the effect of the X-ray heating is significant.
Two-layer flows are expected in this case too but the flow rates through the thin disks are largely dominant to those through the thick flows.

The typical $\dot{M}_{0}$ in the present study is estimated in equation (\ref{eqn:Mdot_0}) and it shows that $\dot{M}_{0}$ is around $\dot{M}_{2}$ or more if we consider cases of $M \sim 10^{7} \sim 10^{8} M_{\odot}$.
In that $\dot{M}_{0}$ range, the effect of X-ray heating on 
the envelope is significant and we can expect a thermally unstable situation toward a two-phase equilibrium studied by Krolik, McKee and Tarter (1981):
If a certain small region in the envelope in a thermal balance between radiative cooling and X-ray heating increases its density above the average, the matter in that region starts cooling due to enhanced radiative cooling, and shrinking since the pressure gets less than the ambient one, getting denser and denser.  On the other hand, if another region in the envelope decreases its density, the matter in that region starts heating and expanding due to enhanced X-ray heating.
As a result, the matter in the envelope is expected to establish a two phase system in which a number of dense and cold clouds move around in a hot and extended inter-cloud gas.
Those cold clouds are thought to sink towards the envelope center as the cooling flow.

This situation can explain why there exist cold clouds  emitting optical broad lines in the broad line region.  
It can also interpret recent observations of rapid variabilities in absorption of X-rays from several AGNs by considering that the X-ray source is partially obscured by warm or cold clumps which are moving in the broad line region (see e.g. section 7 in Inoue 2021c for a brief review and references therein).

\subsection{Thin excretion disk}
Inoue (2021b) predicts that a thin excretion disk extends from the accretion ring but it terminates at a radius, $r_{\rm t}$, with distance 4 times the accretion ring radius, $r_{\rm k}$, forming another ring there.
Namely,
\begin{equation}
r_{\rm t} = 4 r_{\rm k}.
\label{eqn:r_t-r_k}
\end{equation}

In the case of X-ray binaries, the termination distance of the thin excretion disk exceeds the Roche lobe radius and the tides from the companion star could prevent the disk extension to the termination radius.
In AGNs, however, no such companion object exists and the extension of the disk to the termination radius is expected to realize.

The structures of the thin excretion flow are considered to be the same as those of the standard accretion disk except that the flow direction is opposite and that the angular momentum outward flow rate is much larger than that inward rate in the accretion disk.
The centrifugal force basically balances with the gravitational force in the radial direction, 
but gets slightly larger than the gravitational force as for the matter to gradually go outward, by the angular momentum transfer from the inner side via viscous stress.

The viscous stress, however, decreases as $r$ increases and gets zero at the termination
radius.
Thus, the energy generation rate through the viscous stress decreases and the disk temperature decreases towards the termination radius.
As a result, the outflowing matter through the thin excretion disk is expected to accumulate at the termination radius, forming an excretion ring with a very low temperature.

This reminds us of an observational evidence that the innermost radii of the dust torus observed in infrared rays have a proportionality to square root of the optical luminosity similarly to the broad line region and are 4 to 5 times the radii of the broad line region (e.g. Koshida et al. 2014).
Although the proportionality is currently explained by the dust sublimation model (Barvainis 1987), the excretion ring could be another possibility to interpret the properties of the innermost radii of the dust torus.

The matter flowing out through the excretion disk finally accumulates in the excretion ring, and the radius of the ring is determined by the specific angular momentum carried by the inflowing matter from the tail shock region.
If the specific angular momentum is really given by the density gradient and/or the Coriolis force as discussed in subsection \ref{accretion_ring}, the amplitude and direction of the specific angular momentum vector could possibly vary depending on the position of the black hole in the circum-nuclear region, and then the radius and the rotational axis of the excretion ring is likely to change in time.
A locus of an excretion ring which have been changing its radius and axis could be the origin of thin water maser disks sometimes exhibiting warps observed from some AGNs (see e.g. Herrnstein et al. 1996; Greenhill et al. 2003).

\subsection{Thick outward flow}
Inoue (2021b) also discusses that a thick excretion flow spreads out from the accretion ring.

According to Inoue (2021b), the equation for the angular momentum transfer in the thick excretion flow  is approximately expressed by integrating the physical quantities over the direction perpendicular to the equatorial plane  as
\begin{equation}
\dot{M}_{\rm tk,\; ex} \ell_{\rm ex} - 2\pi r^{2} W_{r\phi, \; ex} \simeq 2 \dot{M}_{\rm tk,\; ex} \sqrt{r_{\rm k}GM},
\label{eqn:thickAMoutflow_rate}
\end{equation}
where $\ell_{\rm ex}$ is the specific angular momentum of the outflowing matter and $W_{\rm r\phi,\; ex}$ is the vertically integrated viscous stress.
The equation for the energy flow rate is written as 
\begin{equation}
\dot{M}_{\rm tk,\; ex} \left(\frac{v^{2}}{2} + \frac{5kT}{m_{\rm p}} - \frac{GM}{r}\right)  - 2\pi r^{2} W_{r\phi, \; ex} \Omega \simeq \dot{M}_{\rm tk,\; ex} \frac{GM}{r_{0}},
\label{eqn:Energy_Eq}
\end{equation}
where $v$, $T$ and $\Omega$ are the velocity, the temperature and the angular velocity of the outflowing matter and we assume $\varepsilon_{0} = 0$ in equation  (\ref{eqn_EnEq-r_k}).  

We approximate the specific angular momentum, $\ell_{\rm ex}$, from equation (\ref{eqn:thickAMoutflow_rate}) as
\begin{equation}
\ell_{\rm ex} \simeq \left\{
\begin{array}{ll}
\sqrt{rGM} & \mbox{  when }r_{\rm t} \ge r \ge r_{\rm k} \\
\sqrt{r_{\rm t}GM} & \mbox{  when }r > r_{\rm t} .
\end{array}
\right.
\label{eqn:j_thick}
\end{equation}
Then, equation (\ref{eqn:Energy_Eq}) can be rewritten for the region of $r_{\rm t} \ge r \ge r_{\rm k}$ as
\begin{equation}
\frac{v_{\rm r}^{2}}{2} + \frac{v_{\phi}^{2}}{2} + \frac{5kT}{m_{\rm p}} - \frac{GM}{r} +\frac{GM}{r} \left( 2 \sqrt{\frac{r_{\rm k}}{r}} - 1\right)   \simeq \frac{GM}{r_{\rm k}},
\label{eqn:Energy_Eq_r<r_t}
\end{equation}
where $v_{\rm r}$ and $v_{\phi}$ are the radial and rotational velocity of the outflowing matter.
In this region, it is roughly considered that the rotational velocity is kept Keplerian circular velocity as $v_{\phi} \sim \sqrt{GM/r}$ via the viscous stress and the half thickness of the flow, $h$, is regulated by the pressure gradient as $h \lesssim r$, whereas the radial velocity gradually grows.
In the region of $r > r_{\rm t}$, however, the outflowing matter becomes free from the viscous stress and turns to be a super-sonic flow with the specific energy given as
\begin{equation}
\frac{v^{2}}{2} - \frac{GM}{r} \simeq \frac{GM}{r_{k}},
\label{eqn:EnergyEq_r>r_t}
\end{equation}
and with the terminal velocity at infinity, $v_{\rm tk,\; \infty}$ as
\begin{eqnarray}
v_{\rm tk,\; \infty} &\simeq& \sqrt{\frac{2GM}{r_{\rm k}}} \nonumber \\ 
&\simeq& \frac{ v_{0}}{\sqrt{2} \delta} \nonumber \\
&=& 2.2 \times 10^{8} \left( \frac{v_{0}}{10^{7} \mbox{ cm s}^{-1}}\right) \left( \frac{\delta}{10^{-1.5}} \right)^{-1} \mbox{ cm s}^{-1}.
\label{eqn:v_tk.infty}
\end{eqnarray}

\subsubsection{Formation of bi-polar cones}
The matter in the thick excretion flow is considered to basically rotate circularly above and below the thin excretion disk in the region $r < r_{\rm t}$ but to gradually increase the outward flow-speed as $r$ increases.
The accretion stream, on the other hand, flows towards the accretion ring from the tail shock region.
Since the accretion stream should have a certain thickness, it is expected to interact with the outward flow near the equatorial plane.

Let us compare a ram pressure of the thick excretion flow with that of the accretion stream here.
The mass flow rate the thick excretion flow, $\dot{M}_{\rm tk,\; ex}$, is expressed as
\begin{equation}
\dot{M}_{\rm tk,\; ex} \simeq 4\pi r h n_{\rm tk} m_{\rm p} v_{\rm tk,\; r},
\label{eqn:Mdot_tk}
\end{equation}
where $h$, $n_{\rm tk}$ and $v_{\rm tk,\; r}$ are a half-thickness, a number density and a radial velocity of the outflowing matter.
Then, the ram pressure of the excretion flow, $P_{\rm tk}$, in the radial direction is roughly as
\begin{eqnarray}
P_{\rm tk} &\simeq& n_{\rm tk} m_{\rm p} v_{\rm tk,\; r}^{2} \nonumber \\
&\simeq& \frac{\dot{M}_{\rm tk,\; ex}}{4\pi r h} v_{\rm tk,\; r},
\label{eqn:P_tk}
\end{eqnarray}
where $v_{\rm tk}$ is the total velocity of the outflowing matter.
The mass flow rate through the accretion stream is, on the other hand, given in equation (\ref{eqn:Mdot_st}).
Then, the ram pressure of the accretion stream, $P_{\rm st}$, is roughly estimated as
\begin{equation}
P_{\rm st} \simeq \frac{\dot{M}_{0}}{\pi d^{2}} v_{\rm st}.
\label{eqn:P_st}
\end{equation}
The ratio of $P_{\rm tk}$ to $P_{\rm st}$ is calculated from equations (\ref{eqn:P_tk}) and (\ref{eqn:P_st}) as
\begin{equation}
\frac{P_{\rm tk}}{P_{\rm st}} \simeq \frac{\dot{M}_{\rm tk,\; ex}}{\dot{M}_{0}} \frac{d^{2}}{4 r h} \frac{v_{\rm tk,\; r}}{v_{\rm st}}.
\label{P_tk-P_st}
\end{equation}
It is considered that $d < h < r$ and $\dot{M}_{\rm tk,\; ex} < \dot{M}_{0}$.
Hence, we could say that the ram pressure of the excretion flow is well less than that of the accretion stream.

The above considerations induce an inference that the thick outward flow could ride up on the accretion stream and get an upward momentum from the equatorial plane by a repulsive force from the accretion stream. 
This could make each of the bottom boundary surfaces of the outward flow above and below the thin excretion disk a hollow cone with a fairly large opening angle, forming bi-polar cones.
Figure \ref{fig:ExcretionFlow} schematically exhibits the flow direction changes of the thick excretion flow due to the interaction with the accretion stream.

\subsubsection{Thermal situation of the outward flow}
Thermal situation of the matter in the thick excretion flow can be judged by the ionization parameter, $\Xi$, defined to be $F/(n_{\rm tk} kT_{\rm tk} c)$ by Krolik, McKee \& Tarter (1981).
$T_{\rm tk}$ is the temperature of the excretion flow and $F$ is a bolometric flux illuminating the matter which is approximated as $F \simeq L/(4\pi r^{2})$.
Then, $\Xi$ is rewritten to be $L/(4\pi n_{\rm tk} r^{2} kT_{\rm tk} c)$.
$n_{\rm tk}r^{2}$ can be given from equation (\ref{eqn:Mdot_tk})  as
\begin{equation}
n_{\rm tk}r^{2} \simeq \frac{\dot{M}_{\rm tk,\; ex}}{4\pi m_{\rm p} v_{\rm tk,\; r}} \left( \frac{h}{r}\right)^{-1}
\label{eqn_nr^2}
\end{equation}
in the region of $r_{\rm k} \le r \le r_{\rm t}$, while 
$L \simeq (\dot{M}_{0}/2)  \eta c^{2}$ as given in equation (\ref{eqn:L_X}).
Thus, we can calculate the ionization parameter, $\Xi$, of the thick excretion flow in the above region as
\begin{eqnarray}
\Xi &\simeq& \frac{\dot{M}_{0} \eta c h v_{\rm tk,\; r}}{2 \dot{M}_{\rm tk,\; ex} r kT_{\rm tk}} \nonumber \\
&\simeq& \frac{\eta c v_{\rm tk,\infty}}{kT_{\rm tk}} \frac{\dot{M}_{0}/2}{\dot{M}_{\rm tk,\; ex}} \frac{v_{\rm tk,\; r}}{v_{\rm tk, \infty}} \frac{h}{r} \nonumber \\
&\simeq& \frac{ \eta m_{\rm p} v_{0} c}{\sqrt{2} \delta k T_{\rm tk}}  \frac{\dot{M}_{0}/2}{\dot{M}_{\rm tk,\; ex}} \frac{v_{\rm tk,\; r}}{v_{\rm tk, \infty}} \frac{h}{r}
\label{eqn:Xi}
\end{eqnarray}
with the help of equation (\ref{eqn:v_tk.infty}).
By employing the approximation of the standard accretion disk theory (Shakura \& Sunaev 1973), $h/r$ can be expressed as
\begin{eqnarray}
\frac{h}{r} &\simeq& \left( \frac{kT_{\rm tk}r_{\rm k}}{m_{\rm p}GM}\right)^{1/2} \left( \frac{r}{r_{\rm k}}\right)^{1/2}  \nonumber \\ 
&\simeq& 3.1 \times 10^{-1} \left( \frac{T_{\rm ik}}{3.2 \times 10^{7} \mbox{ K}}\right)^{1/2} \left( \frac{\delta}{10^{-1.5}}\right) \left( \frac{v_{0}}{10^{7} \mbox{ cm s}^{-1}}\right)^{-1}  \left( \frac{r}{r_{\rm k}}\right)^{1/2}.
\label{eqn:h/r}
\end{eqnarray}
If we consider that the matter with the temperature, $T_{\rm tk} = T_{\rm in} = 3.2 \times 10^{7}$ K, is flowing from the accretion ring into the excretion flow, and adopt $\eta= 0.1$, $\delta = 10^{-1.5}$ and $v_{0} = 10^{7}$ cm s$^{-1}$, and $v_{\rm tk, \; r} = 0.1 v_{\rm tk,\; \infty}$ as  reference values, equation (\ref{eqn:Xi}) can be estimated as
\begin{equation}
\Xi \simeq 6.6  \left( \frac{\dot{M}_{\rm tk,\; ex}}{\dot{M}_{0}/2} \right)^{-1} \left( \frac{h/r}{0.31} \right) \left( \frac{v_{\rm tk,\; r}/v_{\rm tk,\; \infty}}{0.1}\right).
\label{eqn:Xi-value}
\end{equation}

Krolik, Mckee \& Tarter (1981) show that a situation in which a significant fraction of the matter is condensed into a number of cold clouds as discussed in 
subsection \ref{accretion_ring} is possible to appear when $\Xi$ is around unity. 
The value of $\Xi$ for the reference parameter values in the above equation is already close to the unstable range.
Furthermore, if the thick outward flow rides up on the stream as discussed above, it could cause a compression of the flow thickness and a brake of the outward velocity.
Then, if the product of the decreases of $h/r$ and $(v_{\rm tk,\; r}/v_{\rm tk, \; \infty})$ gets smaller by a factor several in equation (\ref{eqn:Xi-value}), $\Xi$ is possible to become unity. 
In that case, cool and dense clouds are expected to appear in the excretion flow similarly to the case in the accretion ring.
Differently from the case of  the accretion ring, however,
the emergent cold clouds should have enough kinetic energy to escape to the infinity in the excretion flow, while those in the accretion ring should sink into the envelope center. 
Hence, the cold clouds appearing in the excretion flow are expected to flow outwards along the outer edge of the bi-polar cones.
This could interpret the mid-infrared polar cones observed from several AGNs (see e.g. H\"{o}nig 2019 and references therein).

These cold clouds could also play X-ray absorbers.
The average column density of the thick excretion flow in the radial direction near the accretion ring is roughly estimated from equation (\ref{eqn:Mdot_tk}) as 
\begin{eqnarray}
n_{\rm tk} r &\simeq& \frac{\dot{M}_{\rm tk,\; ex}}{4\pi m_{\rm p} h v_{\rm tk,\; r}} \nonumber \\
&\simeq& \frac{ v_{0}}{\sqrt{8} \delta \eta c \sigma_{\rm T}} \left(\frac{\dot{M}_{\rm tk,\; ex}}{\dot{M}_{0}/2}\right) \left(\frac{\dot{M}_{0}}{\dot{M}_{\rm E}}\right) \left(\frac{v_{\rm tk,\; r}}{v_{\rm tk, \infty}}\right)^{-1} \left(\frac{h}{r}\right)^{-1} \left(\frac{r}{r_{\rm k}}\right)^{-1} \nonumber \\
&\simeq& 1.9 \times 10^{23} \left(\frac{\delta}{0.1}\right)^{-1} \left(\frac{\eta}{0.1}\right)^{-1} \left(\frac{v_{0}}{10^{7} \mbox{ cm s}^{-1}}\right) \left(\frac{\dot{M}_{\rm tk,ex}}{\dot{M}_{0}/2}\right) \left(\frac{\dot{M}_{0}/\dot{M}_{\rm E}}{0.1}\right)  \nonumber \\ && \left(\frac{v_{\rm tk,\; r}/v_{\rm tk, \infty}}{0.1}\right)^{-1} \left(\frac{h/r}{0.31}\right)^{-1} \left(\frac{r}{r_{\rm k}}\right)^{-1} \mbox{ cm}^{-2}
\label{eqn:n_tk-r}
\end{eqnarray}
with the help of equations (\ref{eqn:r_k}) and (\ref{eqn:Mdot_E}).
If a fairly large fraction of the matter in the excretion flow is condensed into cold clouds, the column density integrated over the clouds on the line of sight can be expected to be sufficiently large for significant X-ray absorption unless the position, $r$, gets much larger than $r_{\rm k}$ and the velocity, $v_{\rm tk,\; r}$, becomes close to $v_{\rm tk, \infty}$.
The above discussions indicate that the dense and cold clouds in the region of the thick excretion flow where the collimation takes place could largely contribute to X-ray absorptions often seen in the type 2 AGNs.
This indication looks consistent with the unified picture drawn from X-ray observations (see e.g. Ogawa et al. 2021).

It should be noted then that $\Xi$ in equation (\ref{eqn:Xi-value}) has a term of $[\dot{M}_{\rm tk,\; o}/(\dot{M}_{0}/2)]^{-1} = (\dot{M}_{\rm tk}/\dot{M}_{0})^{-1}$. As briefly mentioned in subsection \ref{accretion_ring}, Inoue (2021b) shows that the ratio of $\dot{M}_{\rm tk}/\dot{M}_{0}$ decreases from $\sim 1$ to $\sim 0.1$ or less as $\dot{M}_{0}$ increases from a small value less than $\dot{M}_{1}$ in equation (\ref{eqn:M_1}) to $\dot{M}_{2}$ close to the Eddington accretion rate in equation (\ref{eqn:M_2}).
Hence, a degree for the unstable situation with $\Xi \sim 1$ to appear in the thick outward flow is expected to decrease as $\dot{M}_{0}$ increases over the above range.
This could be able to interpret the relation between the fraction of obscured AGN and the Eddington ratio obtained by Ricci et al. (2017).

\subsection{Interaction of the ISM flow with the bi-polar cones}
The bi-polar outward flow should interact with the ISM flow towards the tail shock region behind the black hole.
The ram pressure in the radial direction of the polar outward flow, $P_{\rm po}$, is approximately calculated as
\begin{equation}
P_{\rm po} \simeq \frac{\dot{M}_{\rm tk,\; ex} v_{\rm tk,\; r}}{\Omega r^{2}},
\label{eqn:P_pe}
\end{equation}
where $\Omega$ is a total solid angle of the two hollow polar cones as viewed from the center.
That in the flow direction of the interstellar matter, $P_{\rm is}$, is roughly given as
\begin{equation}
P_{\rm is} \simeq \chi \frac{\dot{M}_{0}v_{0}}{\pi r_{\rm a}^{2}},
\label{eqn:P_is}
\end{equation}
where $\chi$ is a enhancement factor due to a shrinkage of the cross section of the hyperbolic flow towards the black hole and could be a factor of a few.
We can calculate a ratio of $P_{\rm is}$ to $P_{\rm po}$ from the above two equations as
\begin{equation}
\frac{P_{\rm is}}{P_{\rm po}} \simeq \chi \frac{\Omega}{\pi} \frac{\dot{M}_{0}}{\dot{M}_{\rm tk,\; ex}} \frac{v_{0}}{v_{\rm tk,\; r}} \left(\frac{r}{r_{\rm a}}\right)^{2}.
\label{eqn:P_is-P_pe}
\end{equation}
Considering $\chi \sim 2 \sim 3$, $\Omega \sim \pi$,  $v_{\rm tk,\; r} \sim 10 v_{0}$ and $\dot{M}_{0}$ is  $\sim 10 \dot{M}_{\rm tk,\; ex}$ at most when $\dot{M}_{0} \gtrsim \dot{M}_{2}$, we see that $P_{\rm is}$ can be comparable to $P_{\rm po}$ in a region of $r \sim r_{\rm a}$ but that $P_{\rm is}$ is less than $P_{\rm po}$ where $r$ is well less than $r_{\rm a}$.
When $P_{\rm is}$ is comparable to $P_{\rm po}$, the boundary of the polar cone on an interstellar matter approaching side could possibly be pushed inward.
Hence, the above expected relation between $P_{\rm is}$ and $P_{\rm po}$ depending on $r$ could explain the narrowing of the opening angle of the polar cone from $\sim 120^{\circ}$ in the sub-pc region to $\sim 90^{\circ}$ in the region of $r >$ a few pc observed from the Circinus Galaxy (Tristram et al. 2007).

Even though the opening angle of the polar cone is narrowed in the region of $r \gtrsim r_{\rm a}$, the  ISM flow is, as a whole, considered to be forced to get under the hollow polar cones.
In the region of $r \ll r_{\rm a}$, in particular, the ISM flow is expected to be constrained in a disk like region sandwiched by the two polar cones with the wide opening angle.
This could correspond to the mid-infrared disk with $\sim$1 pc size observed with the mid-infrared interferometry from NGC 1068 (Raban et al. 2009) and the Circinus Galaxy (Tristram et al. 2014).

The average number density of the interstellar matter in the circum-nuclear region could be $\sim 10^{2}$ cm$^{-3}$ but it could be compressed to be as dense as $\sim 10^{3}$ cm$^{-3}$ in the disk-like region.  
Considering  the size of the region is $\sim 1$ pc, the average column density could be $10^{21.5}$ cm$^{-2}$.
Although the column density should fluctuate since the interstellar matter should be  clumpy, this disk-like region in the ISM flow could not be a heavy X-ray absorber on average.


\section{Summary and discussion}
Figure \ref{fig:FlowChart} shows a block diagram of elements studies in this paper,  and a sequence of the accretion flow from the circum-nuclear region to the black hole and another sequence of the excretion flow from the accretion ring returning to the circum-stellar region. 
This figure also indicates candidate places for the elements required from observations.

First of all, the present study presents the scenario to cause AGN activities by introducing the Bondi-Hoyle-Lyttleton type accretion of interstellar matter by the black hole wandering in a circum-nuclear region.
The accretion rate estimated from the BHL accretion roughly explain the observed luminosities of AGNs.

Wandering of the super-massive black hole in the galactic nucleus region is the starting point of the scenario.
This is based on the recent theoretical study of the wandering mechanism by Inoue (2021a).
This could be supported also by the recent observational study by Combes et al. (2019), who point out that the AGN positions are frequently 
off-centered by several tens pc from the center of the circumnuclear structures 
with the $\sim$ 100 pc size in the CO emission maps of several AGNs observed with ALMA.

If a black hole moves in interstellar medium in a circum-nuclear region, the BHL type accretion is expected to happen.
A number of theoretical works have, however, shown unstable situations of the BHL accretion flows.
They could be divided into two groups in terms of the basic mechanism for the instability.

The first mechanism could be Rayleigh-Taylor and/or Kelvin-Helmholtz instabilities around the bow shock in front of the black hole (see Foglizzo \& Ruffert 1999).
Since the matter accreted by the black hole is mostly from the rear side, however, the large density modulation excited by the instabilities on the front side should propagate to the rear side before the accretion rate is modulated, and its modulation-amplitude could be weakened by the large ratio between the mass incident to the front side and that to the rear side.  In fact, the 3-D simulations by Blondin and Raymer (2012) showed that the amplitude of the resulted accretion rate modulation is moderate.
Furthermore, the time scale of radiative cooling behind the shock is much shorter than the propagation time scale $\sim$ the mass flow time around there, as roughly estimated in subsection \ref{TailShockRegion}, and thus the power of 
the initial density modulation, even if it exists, could be largely damped in course of the propagation.

The second mechanism could be the instability associated with the ionization front (Sugimura \& Ricotti 2020; Newman \& Axford 1967) and it 
should appear under the effects of the radiation from the central engine.
However, we consider a situation here in which the environment around the central engine is separated into two regions: one consisting of two polar cones with fairly wide opening angle in which both radiation and matter are out-flowing from the central engine, and the other, a diks-like region sandwiched by the two polar cones, in which the BHL accretion takes place and the radiation from the central engine is mostly blocked by the matter at the edge of the polar cones.
In this situation, we can expect that the instability due to the radiation effects is highly suppressed too.

The separation into the two regions is consistent with the unified scheme of AGNs (e.g. Antonucci 1993) and the results of recent observations with IR and submillimeter interferometers (e.g. H\"{o}nig 2019).

The key element for the separation of the flows is the accretion ring.

The accretion ring is a place where matter inflowing through the accretion stream from the tail shock region rotates along the Keplerian circular orbit determined by the intrinsic specific angular momentum and sojourns for a while to bifurcate to accretion and excretion flows due to angular momentum transfer in it. 
The specific angular momentum is phenomenologically estimated here by considering that the broad line region often observed in AGNs corresponds to the accretion ring.  Two possibilities for the origin of the angular momentum are discussed only qualitatively.

Inoue (2021b) studies the properties of the accretion ring and predicts that it extends a thick excretion flow and a thin excretion disk outward, as well as a thick accretion flow and a thin accretion disk inward to the black hole.

Recent observations gradually resolve fine structures in environments of AGNs. 
Two components: an equatorial disk/torus and a polar component, are clarified from IR interferometry observations, and presences of H$_{2}$O maser disks are revealed by sub-mm-line interferometry observations (see e.g. Ramos \& Ricci 2017; H\"{o}nig 2019).
In the X-ray regime, the obscuration is shown to be produced by absorbers mostly associated with the ``torus" and the broad-line region (see e.g. Ramos Almeida \& Ricci 2017; Ogawa et al. 2021).
As discussed in the previous section, the structures predicted for the excretion flow and disk in the present scenario can provide possible origins of such substructures. 

As a whole, the present study presents a likely guide line to understand the overall accretion environments of AGNs.
The quantitative discussions done above are, however, based only on order estimations and have fairly large ambiguities.
The inferences on the collimation mechanism of the bi-polar cone are primitive too.
Further studies are obviously desired.



\appendix 
\section*{Angular momentum- and energy- transfers from accretion flows to excretion flows}
In this study, it is considered that matter flowing through the accretion stream from the tail shock region first separates to a mass-spreading flow and a cooling flow in the ring-envelope and then each of the flows further bifurcates to an accretion flow and an excretion flow.  
As a result, two pairs of flows are expected to appear: a thick accretion flow and a thick excretion flow, and  a thin accretion disk and a thin excretion disk.

For the pair of the thick flows, equations of the angular momentum transfer are given as 
\begin{equation}
-\dot{M}_{\rm tk, \; ac} \ell_{\rm ac} - 2\pi r^{2} W_{r\phi, \; \rm ac} = -\dot{M}_{\rm tk,\; ac} \ell_{\rm in},
\label{eqn:AMT_tk_ac}
\end{equation}
for the thick accretion flow, and
\begin{equation}
\dot{M}_{\rm tk, \; ex} \ell_{\rm ex} - 2\pi r^{2} W_{r\phi, \; \rm ex} = \dot{M}_{\rm tk,\; ex} \ell_{\rm out},
\label{eqn:AMT_tk_ex}
\end{equation}
for the thick excretion flow.
Here, $\ell_{\rm ac}$ and $\ell_{\rm ex}$ are the specific angular momenta of the accretion and excretion matter at a position, $r$, respectively, and $W_{r\phi, \; \rm ac}$ and $W_{r\phi, \; \rm ex}$ are the respective viscous stress integrated over the direction perpendicular to the equatorial plane.
$\ell_{\rm in}$ is the specific angular momentum carried by the accretion matter through the innermost boundary, and $\dot{M}_{\rm tk,\; ac} \ell_{\rm in}$ gives the steady angular-momentum-inflow rate.
On the other hand, $\dot{M}_{\rm tk,\; ex} \ell_{\rm out}$ expresses the angular-momentum-outflow rate, and $\ell_{\rm out}$, the total specific angular momentum transferred outward, is determined at the boundary ($r = r_{\rm k}$) between the accretion flow and the excretion flow, as follows.

Since the specific angular momentum carried by the matter inflowing through the accretion stream to the boundary at $r_{\rm k}$ is $\ell_{\rm a}$ as introduced in equation (\ref{eqn:ell_a}), the specific angular momenta of the accretion and excretion matter at the boundary, $\ell_{\rm ac, 0}$ and $\ell_{\rm ex, 0}$, should satisfy the relations as
\begin{equation}
\ell_{\rm ac, 0} = \ell_{\rm ex, 0} = \ell_{\rm a} = \sqrt{r_{\rm k}GM}.
\label{eqn:ell_0}
\end{equation}
Then, the integrated viscous stress of the accretion matter at $r_{\rm k}$, $W_{r \phi, \; \rm ac, 0}$, is given as 
\begin{equation}
2\pi r_{\rm k}^{2} W_{r \phi, \; \rm ac, 0} = - \dot{M}_{\rm tk,\; ac} (\ell_{\rm a} - \ell_{\rm in})
\label{eqn:W_rphi_ac_0}
\end{equation}
from equation (\ref{eqn:AMT_tk_ac}).
Since the integrated viscous stress should continuously act across the boundary, 
the integrated viscous stress of the excretion matter at the inner boundary, $W_{r \phi, \; \rm ex, 0}$,  should relate to $W_{r \phi, \; \rm ac, 0}$ as 
\begin{equation}
W_{r \phi, \; \rm ex, 0} = W_{r \phi, \; \rm ac, 0}.
\label{eqn:TwoW_rphi}
\end{equation}
Thus, the total outflow rate of the angular momentum is calculated from equation (\ref{eqn:AMT_tk_ex}) at $r_{\rm k}$ with the helps of equations (\ref{eqn:ell_0}), (\ref{eqn:W_rphi_ac_0}) and (\ref{eqn:TwoW_rphi}) as
\begin{equation}
\dot{M}_{\rm tk,\; ex} \ell_{\rm out} = \dot{M}_{\rm tk,\; ex} \ell_{\rm a} + \dot{M}_{\rm tk,\; ac} (\ell_{\rm a} - \ell_{\rm in}).
\label{eqn:AMTotalOutflowRate}
\end{equation}
Since $\ell_{\rm in}$ should be much smaller than $\ell_{\rm a}$, we have
\begin{equation}
\ell_{\rm out} \simeq \frac{\dot{M}_{\rm tk,\; ac} + \dot{M}_{\rm tk,\; ex}}{\dot{M}_{\rm tk,\; ex}} \ \ell_{\rm a}.
\label{eqn:ell_out}
\end{equation}

Equations of the energy transfer in the thick accretion and excretion flows are respectively given as
\begin{eqnarray}
&& \dot{M}_{\rm tk,\; ac}\left(\frac{v^{2}}{2} + w -\frac{GM}{r}\right)_{\rm tk, \; ac} + 2\pi r^{2} W_{\rm r\phi,\; \rm ac}\Omega = \nonumber \\ 
&& \dot{M}_{\rm tk,\; ac}\left(\frac{v^{2}}{2} + w -\frac{GM}{r}\right)_{\rm tk, \; ac} - \dot{M}_{\rm tk,\; ac} (\ell_{\rm ac} - \ell_{\rm in})\Omega = \dot{M}_{\rm tk,\; ac} \varepsilon_{\rm tk,\; ac}
\label{eqn:EnTotalInflowRate}
\end{eqnarray}
for the thick accretion flow, and
\begin{eqnarray}
&& \dot{M}_{\rm tk,\; ex}\left(\frac{v^{2}}{2} + w -\frac{GM}{r}\right)_{\rm tk, \; ex} - 2\pi r^{2} W_{\rm r\phi,\; \rm ex} \Omega = \nonumber \\ 
&& \dot{M}_{\rm tk,\; ex}\left(\frac{v^{2}}{2} + w -\frac{GM}{r}\right)_{\rm tk, \; ex} + \dot{M}_{\rm tk,\; ex} (\ell_{\rm out} - \ell_{\rm ex})\Omega = \dot{M}_{\rm tk,\; ex} \varepsilon_{\rm tk,\; ex}
\label{eqn:EnTotalInflowRate}
\end{eqnarray}
for the thick excretion flow.
Here, $v$, $w$ and $\Omega$ are the velocity, the specific enthalpy and the angular velocity of the flowing matter.

$\varepsilon_{\rm tk,\; ac}$ is the specific energy of the accretion matter absorbed by the black hole.
Its value is determined from equation (\ref{eqn:EnTotalInflowRate}) at $r_{\rm k}$,
\begin{equation}
\dot{M}_{\rm tk,\; ac}\left(\frac{v^{2}}{2} + w -\frac{GM}{r}\right)_{\rm tk, \; ac, 0} - \dot{M}_{\rm tk,\; ac} (\ell_{\rm a} - \ell_{\rm in})\Omega_{0} = \dot{M}_{\rm tk,\; ac} \varepsilon_{\rm tk,\; ac}.
\label{eqn:EnTotalInflowRate_Value}
\end{equation}
If we assume
\[
\left(\frac{v^{2}}{2} + w -\frac{GM}{r}\right)_{\rm tk, \; ac, 0} \simeq 0, 
\]
$\ell_{\rm in} \ll \ell_{\rm a} = \sqrt{r_{\rm k}GM}$ and $\Omega_{0} =\sqrt{GM/r_{\rm k}^{3}}$, 
we get
\begin{equation}
\varepsilon_{\rm tk,\; ac} \simeq -\frac{GM}{r_{\rm k}}.
\label{eqn:InflowSpecificEnergy}
\end{equation}

$\varepsilon_{\rm tk,\; ex}$ is the total specific energy carried by the excretion matter and is obtained from equation (\ref{eqn:EnTotalInflowRate_Value}) at $r_{\rm k}$ as
\begin{equation}
\dot{M}_{\rm tk,\; ex}\left(\frac{v^{2}}{2} + w -\frac{GM}{r}\right)_{\rm tk, \; ex, 0} - \dot{M}_{\rm tk,\; ex} (\ell_{\rm out} - \ell_{\rm a})\Omega_{0} = \dot{M}_{\rm tk,\; ex} \varepsilon_{\rm tk,\; ex}.
\label{eqn:EnTotalOutflowRate_Value}
\end{equation}
Assuming again
\[
\left(\frac{v^{2}}{2} + w -\frac{GM}{r}\right)_{\rm tk, \; ex, 0} \simeq 0, 
\]
$\ell_{\rm a} = \sqrt{r_{\rm k}GM}$ and $\Omega_{0} =\sqrt{GM/r_{\rm k}^{3}}$, 
we have
\begin{equation}
\varepsilon_{\rm tk,\; ex} \simeq \frac{\dot{M}_{\rm tk,\; ac}}{\dot{M}_{\rm tk,\; ex}} \frac{GM}{r_{\rm k}}.
\label{eqn:OutflowSpecificEnergy}
\end{equation}

We see from equations (\ref{eqn:InflowSpecificEnergy}) and (\ref{eqn:OutflowSpecificEnergy}) that the equation, 
\begin{equation}
\dot{M}_{\rm tk,\; ex} \varepsilon_{\rm tk,\; ex} + \dot{M}_{\rm tk,\; ac} \varepsilon_{\rm tk,\; ac} = 0,
\label{eqn:EnergyConservation}
\end{equation}
is established to ensure the energy conservation.

For the thin disks, the total specific angular momentum, $\ell_{\rm out,\; tn}$, carried by the thin excretion disk is given as
\begin{equation}
\ell_{\rm out, \; tn} \simeq \frac{\dot{M}_{\rm tn,\; ac} + \dot{M}_{\rm tn\; ex}}{\dot{M}_{\rm tn,\; ex}} \ \ell_{\rm a},
\label{eqn:ell_out_tn}
\end{equation}
via similar calculations as done above.
From this equation, the termination radius of the thin excretion disk, $r_{\rm t}$, is estimated as
\begin{equation}
r_{\rm t} \simeq \left(\frac{\dot{M}_{\rm tn,\; ac} + \dot{M}_{\rm tn\; ex}}{\dot{M}_{\rm tn,\; ex}}\right)^{2} r_{\rm k},
\label{eqn:r_t-general}
\end{equation}
in the general form differently from equation (\ref{eqn:r_t-r_k}) for the specific case of $\dot{M}_{\rm tn, \; ex} = \dot{M}_{\rm tn,\; ac}$.

The total specific energy, $\varepsilon_{\rm tn,\; ex}$,  carried by the thin excretion disk is estimated as
\begin{equation}
\varepsilon_{\rm tn,\; ex} = \frac{\dot{M}_{\rm tn,\; ac} + \dot{M}_{\rm tn\; ex}}{\dot{M}_{\rm tn,\; ex}} \sqrt{\frac{r_{\rm k}}{r}} \frac{GM}{r} - \frac{3}{2} \frac{GM}{r},
\label{eqn:OutflowSpecEngy_tn}
\end{equation}
from equation (\ref{eqn:EnTotalInflowRate}) after replacing all the subscripts ``tk" to ``tn" and $\ell_{\rm out}$ to $\ell_{\rm out,\; tn}$ in equation (\ref{eqn:ell_out_tn}), and approximately setting $v = \sqrt{GM/r}$, $w=0$, $\Omega = \sqrt{GM/r^{3}}$, and $\ell_{\rm ex} = \sqrt{rGM}$ on the assumption that the matter basically rotates with the Keplerian circular velocity and the thermal energy is negligibly small in the thin excretion disk, the same as for the thin accretion disk.
Here, $\varepsilon_{\rm tn,\; ex}$ changes with $r$ due to the blackbody emission from the disk surface and its value at the inner boundary, $r_{\rm k}$, is given as
\begin{equation}
\varepsilon_{\rm tn,\; ex, 0} = \frac{\dot{M}_{\rm tn,\; ac} + \dot{M}_{\rm tn\; ex}}{\dot{M}_{\rm tn,\; ex}} \frac{GM}{r_{\rm k}} - \frac{3}{2} \frac{GM}{r_{\rm k}}.
\label{eqn:OutflowSpecEngy_tn_0}
\end{equation}

Equation (\ref{eqn:r_t-general}) shows that $r_{\rm t}$ can get to infinity with the infinitesimally small $\dot{M}_{\rm tn,\; ex}$, and there has been an argument in the literature that such a situation as $\dot{M}_{\rm tn,\; ex} \ll  \dot{M}_{\rm tn,\; ac}$ could arise in the accretion disk (e.g. Pringle 1981).
However, this situation requires an artificially large specific energy at $r_{\rm k}$ as seen from equation (\ref{eqn:OutflowSpecEngy_tn_0}).

Similarly, equation (\ref{eqn:OutflowSpecificEnergy}) tells us that the total specific energy, $\varepsilon_{\rm tk,\; ex}$, carried by the thick excretion flow gets unrealistically large also when  $\dot{M}_{\rm tk,\; ex} \ll  \dot{M}_{\rm tk,\; ac}$.

\end{document}